\documentclass{article}
\usepackage[numbers]{natbib} 

\usepackage[preprint]{neurips_2025}

\usepackage[utf8]{inputenc} 
\usepackage[T1]{fontenc}    
\usepackage{hyperref}       
\usepackage{url}            
\usepackage{booktabs}       
\usepackage{amsfonts}       
\usepackage{nicefrac}       
\usepackage{microtype}      
\usepackage{xcolor}         
\usepackage{graphicx}  
\usepackage{amsmath}
\usepackage{caption}
\usepackage{amsmath}
\usepackage{algorithm}
\usepackage{algpseudocode}
\usepackage{amsmath}
\usepackage{verbatim}
\usepackage{graphicx}

\usepackage{amsmath, amssymb, amsthm}
\usepackage{graphicx}
\usepackage{tikz}
\usetikzlibrary{arrows.meta, positioning}
\usepackage{cleveref}
\usepackage{wrapfig}
\usepackage{enumitem}

\newtheorem{definition}{Definition}
\newtheorem{theorem}{Theorem}
\newtheorem{proposition}{Proposition}

\newcommand{\indep}{\perp \!\!\! \perp}

\definecolor{darkred}{rgb}{0.7, 0.0, 0.0}

\title{Epiphenomenal Abstractions for Causal Effect Estimation}
\title{Causal Formulation of Brain Waves}
\title{Duality of Spike Wave: A Primer Causal Formulation}
\title{Causal Formulation of Spike-Wave Duality}
\title{A Causal Formulation of Spike-Wave Duality}

\author{
  Kasra Jalaldoust\thanks{The authors contributed equally.}\\
  Department of Computer Science\\
  Columbia University\\
  \texttt{kasra@cs.columbia.edu}
  \And
  Erfan Zabeh\footnotemark[1]\,  \thanks{Corresponding author.}\\
  Mortimer B. Zuckerman Mind Brain Behavior Institute\\
  Columbia University\\
  \texttt{erfan.zabeh@columbia.edu}
}

\begin{document}

\maketitle

\begin{abstract}

Understanding the relationship between brain activity and behavior is a central goal of neuroscience. Despite significant advances, a fundamental dichotomy persists: neural activity manifests as both discrete spikes of individual neurons and collective waves of populations. Both neural codes correlate with behavior, yet correlation alone cannot determine whether waves exert a causal influence or merely reflect spiking dynamics without causal efficacy. According to the Causal Hierarchy Theorem, no amount of observational data—however extensive—can settle this question; causal conclusions require explicit structural assumptions or careful experiment designs that directly correspond to the causal effect of interest. We develop a formal framework that makes this limitation precise and constructive. Formalizing epiphenomenality via the invariance of interventional distributions in Structural Causal Models (SCMs), we derive a certificate of sufficiency from Pearl’s do-calculus that specifies when variables can be removed from the model without loss of causal explainability and clarifies how interventions should be interpreted under different causal structures of spike-wave duality. The purpose of this work is not to resolve the spike–wave debate, but to reformulate it. We shift the problem from asking which signal matters most to asking under what conditions any signal can be shown to matter at all. This reframing distinguishes prediction from explanation and offers neuroscience a principled route for deciding when waves belong to mechanism and when they constitute a byproduct of underlying coordination

\end{abstract}

\section*{Introduction and background}

Neuroscience seeks to explain how diverse forms of neural activity give rise to behavior. These forms appear in many layers: the discrete action potentials of \emph{spikes}, the collective oscillations of \emph{waves}, and slower modulatory processes such as neuromodulator release, glial signaling, or hemodynamic changes. Each can be measured, each can be correlated with behavior, but correlation alone does not reveal causal force. A variable may appear essential while in fact being \emph{epiphenomenal}: a secondary effect that mirrors true causes without altering them \cite{Pearl2009, Woodward2003}. Oscillations have long stood at the center of this dilemma \cite{van2025processes, Buzsaki2004}. They are celebrated as carriers of coordination and dismissed as echoes of spiking activity, both at once \cite{Buzsaki2004, Buzsaki2006, Fries2005, Fries2015, Miller2009, Ray2011, Bressler2015}. To call waves epiphenomenal is not a small claim—it suggests that what looks like the music of the brain may in fact be only its resonance. To know the difference requires more than observation; it demands a causal lens that can separate what drives from what follows.

Causal inference has been long developed in computer science and statistics, and has proven its application in fields such as economics and social sciences\cite{Imbens2015, Hernan2020,pearl_causal,hunermund2025causal}. 
In neuroscience, causality has traditionally been interpreted through stimulation and perturbation, though recent work has begun to examine it as a conceptual problem in its own right \cite{kording2006causal, marinescu2018quasi, bailey2024causal, siddiqi2022causal}. Yet, no formal framework has been proposed to analyze the causal structure of brain waves and spikes. This absence underscores the need for a theoretical foundation—complementary to experimental approaches—that clarifies under what conditions waves can be said to exert causal influence on spiking activity and behavior.


\begin{figure}[t]
\centering
\includegraphics[width=0.99\textwidth]{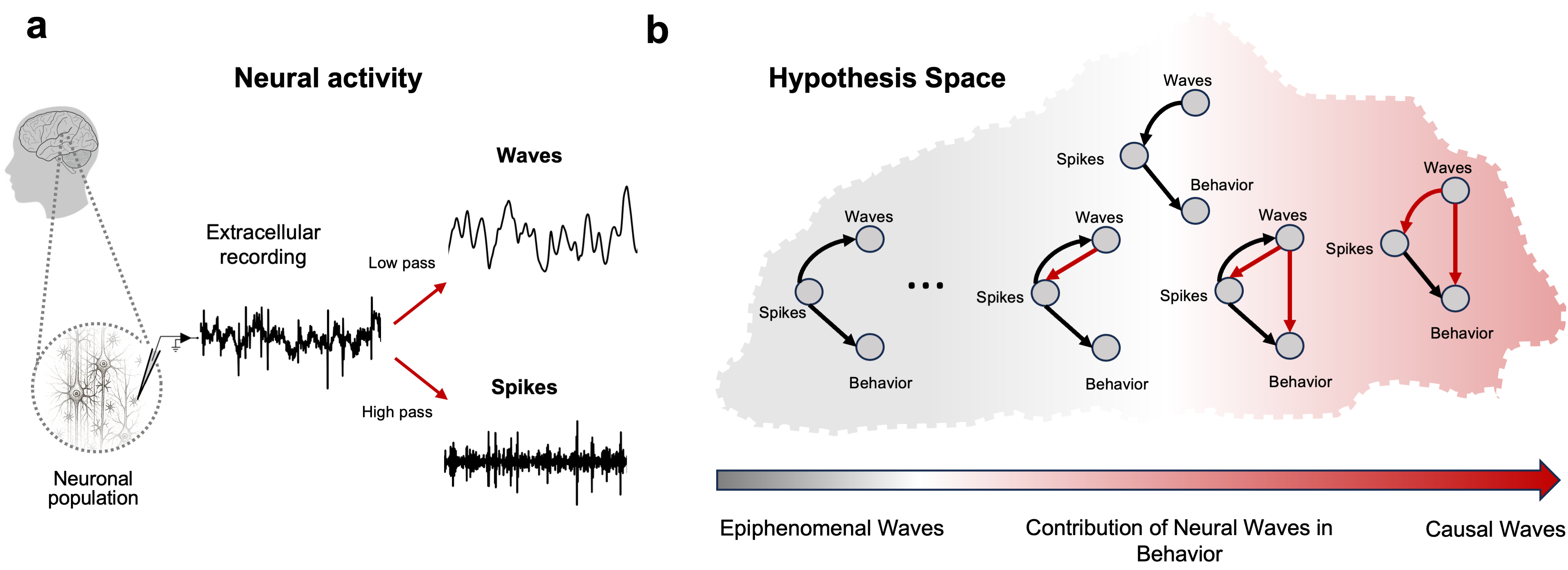}  
\caption{
 \textbf{Neural activity and hypothesis space for spike–wave–behavior interactions.}
\textbf{(a)} Raw neuronal recording can be decomposed into population-level waves (low-pass filtered) and single-neuron spikes (high-pass filtered). These two levels of description coexist within the same signal but may play distinct roles in shaping behavior.
\textbf{(b)} Candidate causal graphs span a spectrum from epiphenomenal to causal waves. Each node represents waves, spikes, or behavior; arrows denote possible causal dependencies. Red edges highlight putative direct influences of waves. Graphs on the left illustrate epiphenomenal waves, where waves carry no causal force once spikes are accounted for.
 }
\label{fig:example}
\end{figure}
The predominant view in neuroscience holds that spiking activity constitutes the fundamental language of computation between neurons \cite{Rieke1999, Dayan2001}. This perspective is grounded in several compelling arguments that suggest neural oscillations may be epiphenomenal rather than causally relevant to behavior. The spatial resolution limitations of field potential recordings raise fundamental questions about the precision with which population-level signals can interact with individual neurons. Neural oscillations are typically measured using relatively low-resolution recording electrodes, which aggregate activity across large populations of neurons \cite{Buzsaki2012, Einevoll2013}. This coarse spatial sampling brings into question how fine-grained these population effects can be when interacting with individual neurons, potentially limiting their capacity for precise computational influence \cite{Kajikawa2011, Linden2011}. Furthermore, the most persistent neuronal oscillations are prominently observed during sleep and unconscious conditions, suggesting their potential irrelevance to active behavioral processes in the brain \cite{Steriade1993, Hobson2002, muller2016rotating,Diekelmann2010}. The prevalence of oscillatory activity during states of reduced consciousness has led some researchers to view these rhythms as byproducts of neural network dynamics rather than active contributors to cognitive function \cite{Tononi2006, Vyazovskiy2013}.

However, extensive evidence supports the view that neural waves may indeed exert causal influence on behavior and neural dynamics. Multiple contributing factors beyond spiking activity, including neuromodulatory effects and subthreshold activity, are known to contribute to neuronal dynamics \cite{Marder2012, Bargmann2012, Destexhe2003, gedankien2025cholinergic}. Field potentials are thought to include components reflecting the aggregated effects of this subthreshold, non-spiking activity, potentially providing a mechanism for causal influence \cite{Haider2006, Okun2008}. The clinical applicability of brain waves provides compelling evidence for their functional relevance. Neural oscillations are successfully used in medical contexts for therapeutic monitoring, such as tracking anesthesia states in surgery rooms through field potential measurements \cite{Purdon2013, Brown2010,choubdar2023neural}. 
This level of clinical applicability reinforces the functional relevance of brain waves beyond their well-documented behavioral associations across cognitive domains such as attention, memory, and navigation \cite{miller2018working,zabeh2023beta,buzsaki2004neuralosc,losonczy2010network,zhang2018theta,lega2012human}.

From a biophysical perspective, even the most conservative view acknowledges that field potentials represent summations of population-level spiking activity and carry information about system states that can interact with individual neurons \cite{Anastassiou2011, Frohlich2010}. The presence of gap junctions in the brain is well-established, and accordingly, the effect of population-level voltage signals on individual neurons through electrical coupling is biophysically plausible \cite{Connors2004, Bennett2004, Hormuzdi2001}. Furthermore, field potentials contain both oscillatory and non-oscillatory components, and much of the debate focuses on the oscillatory component while less attention has been paid to non-periodic field potentials \cite{He2014, Miller2009b, donoghue2020parameterizing}. This distinction is important because different components of the field potential may have different causal roles in neural computation.

Analyses of spike–wave relationships have typically relied on correlation-based approaches, which are inherently limited. According to the Causal Hierarchy Theorem \cite{bareinboim-r60}, making causal claims requires one of two elements beyond observational data: (1) explicit causal assumptions, or (2) interventional data from randomized controlled trials. This fundamental limitation means that no amount of observational data, regardless of sample size, can definitively resolve questions of causality without additional assumptions or experimental interventions. This theoretical constraint is particularly relevant to neuroscience, where perturbations are not necessarily equivalent to causal interventions \cite{Krakauer2017}. The complex, interconnected nature of neural systems means that experimental manipulations may have cascading effects that obscure rather than clarify causal relationships.

This paper addresses these fundamental challenges by proposing a formal causal framework for analyzing spike-wave duality. Rather than attempting to resolve the spike-wave debate directly, our contribution is to provide a causal paradigm for reasoning about epiphenomenality in this context. This framework specifies how one might formally test when wave-behavior interactions are causally sufficient versus epiphenomenal, leaving the ultimate empirical conclusion to the neuroscience community. We present our results through a structured analysis that formalizes the conditions under which neural waves can be considered epiphenomenal or causally relevant, providing both theoretical foundations and practical tools for empirical investigation.

\section*{Problem statement and causal formulation}

\begin{wrapfigure}{r}{0.2\textwidth}
\centering
\includegraphics[width=\linewidth]{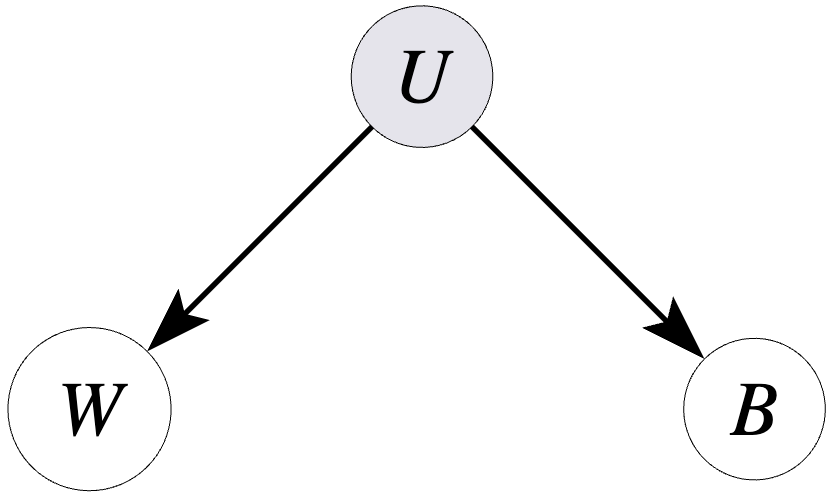}
\caption{Confounded graph}
\label{fig:confounded}
\end{wrapfigure}
We present the problem statment and results in form of a debate between two fictional neuroscientists; John, who believes that waves are epiphenomenal, and Earl, who believes in causal relevance of waves to the behavior. 

Earl has conducted a passive (i.e., non-interventional) brain recording that involves behavior and waves simultaneously. He observes that waves and behavior are statistically dependent, i.e., $P(\mathbf{B}\mid \mathbf{W}) \neq P(\mathbf{B})$. Based on this observation, Earl hypothesizes that waves causally influence the behavior, and presents the findings to John. John disagrees with the causal conclusion, and speculates that the statistical dependence might be solely due to unobserved confounding, without any causal influence from waves to behavior. John's intuition is represented by the \textit{confounded} graph shown in \Cref{fig:confounded}; here, $\mathbf{B}$ obtains it's value as a function of unobserved confounder $\mathbf{U}$, without any mechanistic influence from $\mathbf{W}$. Earl realizes that his passively collected data can not refute the possibility presented by John.

\begin{wrapfigure}{r}{0.2\textwidth}
\centering
\includegraphics[width=\linewidth]{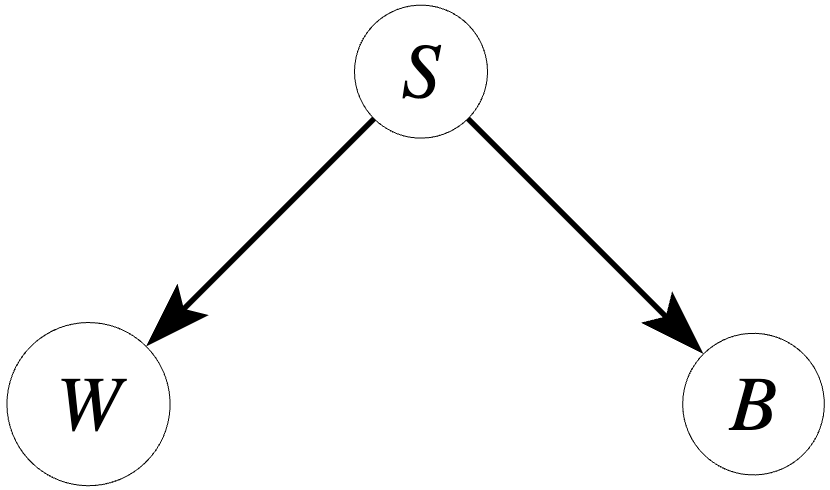}
\caption{Fork graph}
\label{fig:fork}
\end{wrapfigure}

In an attempt to formalize the causal claim, Earl designs an imaginary controlled experiment. He would perturb the waves physically to override the influence from the confounders $\mathbf{U}$ on $\mathbf{W}$, and at the same time samples from behavior under this intervention. The corresponding distribution is denoted by $P(\mathbf{B}; do(\mathbf{W}))$ in causal inference notation. Next, Earl would compare it with the passive/observational distribution of behavior denoted as $P(\mathbf{B})$. This comparison is a sufficient criterion to assess the causal claim:

\begin{enumerate}[leftmargin=1em]
    \item[-] If $P(\mathbf{B}; do(\mathbf{W})) \neq P(\mathbf{B})$, then there exists \textit{some} causal influence from waves on behavior, and more refined measures are needed to quantify this causal effect.
    \item[-] If $P(\mathbf{B}; do(\mathbf{W})) = P(\mathbf{B})$, then there is no causal influence from waves on behavior.
\end{enumerate}


As the debate is unsettled, John contemplates the idea that the spikes generated by certain neurons are the sole determinant of behavior. If true, this deems waves \textit{epiphenomenon}, with no causal relevance to behavior given the spikes are measured; the graph in \Cref{fig:fork} summarizes this narrative. Notably, considering waves alongside spikes offers a better prediction for behavior, thus, simply discarding the waves is unjustified as long as behavior prediction is a scientific objective.\cite{Sani2024, Hsieh2019}. In probabilistic terms we write,

\begin{wrapfigure}{r}{0.2\textwidth}
\centering
\includegraphics[width=\linewidth]{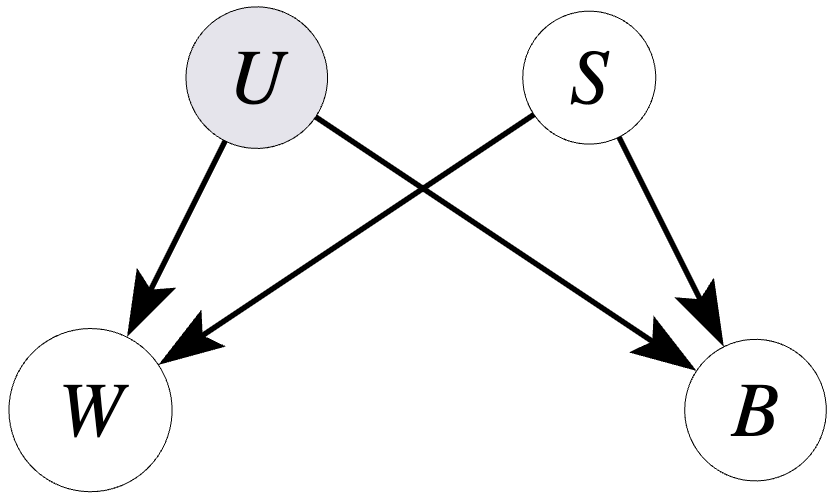}
\caption{Confounded fork graph}
\label{fig:c-fork}
\end{wrapfigure}

\begin{equation} \label{eq:conditional-dependence}
    P(\mathrm{behavior} \mid \mathrm{spikes}, \mathrm{waves}) \neq P(\mathrm{behavior} \mid \mathrm{spikes}).
\end{equation}

Notably, \Cref{eq:conditional-dependence} is inconsistent with the fork graph; rule 1 of do-calculus implies that for all structural causal model that induce the fork graph, we have equality instead of inequality.

If both the fork graph (\Cref{fig:fork}) and confounded graph (\Cref{fig:confounded}) are not consistent with the brain data, what would be compatible graph? A plausible scenario is summarized in \Cref{fig:c-fork}; there are observed and unobserved phenomenon both causing the waves and behavior, simultaneously. John argues that if this graph is true, then the waves are epiphenomenon, since there is no direct causal influence from $\mathbf{W}$ to $\mathbf{B}$.

\begin{wrapfigure}{r}{0.22\textwidth}
\centering
\includegraphics[width=\linewidth]{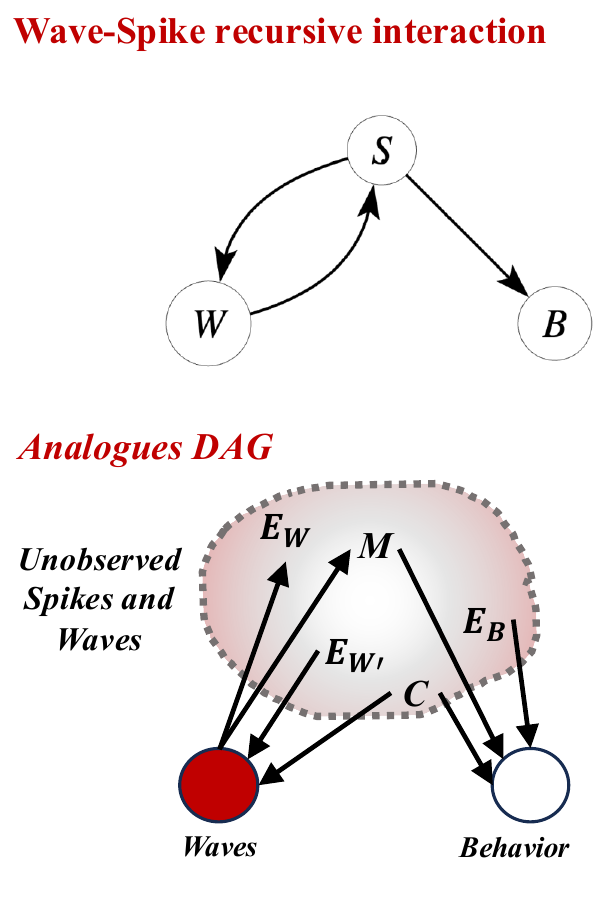}
\caption{Systematic decomposition of unobserved spikes and waves.}
\label{fig:spikes-decomposition}
\end{wrapfigure}

Again, to reject epiphenomenality, Earl must conduct the wave intervention experiment, and to justify epiphenomenality, John must observe all unobserved factors affecting the waves and behavior simultaneously, i.e., absorbing all of $\mathbf{U}$ into $\mathbf{S}$ in the confounded-fork graph.


John proposes that the waves influence some spikes and also get influence by some spikes, without any direct effect on behavior, as shown schematically in \Cref{fig:spikes-decomposition}. To investigate this further, Earl decomposes spikes (and all other factors) into subsets:

\begin{itemize}[noitemsep, topsep=0pt]
    \item \textbf{Confounders $\mathbf{C}$}, such that $\mathbf{C} \to \mathbf{W}$ and $\mathbf{C}\to \mathbf{B}$. 
     \item \textbf{Mediators $\mathbf{M}$}, such that $\mathbf{W} \to \mathbf{M} \to \mathbf{B}$.
     \item \textbf{Exogenous variables $\mathbf{E}$}, such as $\mathbf{E}_W \to \mathbf{W}$ and $\mathbf{E}_B \to \mathbf{B}$.
\end{itemize}

\begin{wrapfigure}{r}{0.2\textwidth}
\centering
\includegraphics[width=\linewidth]{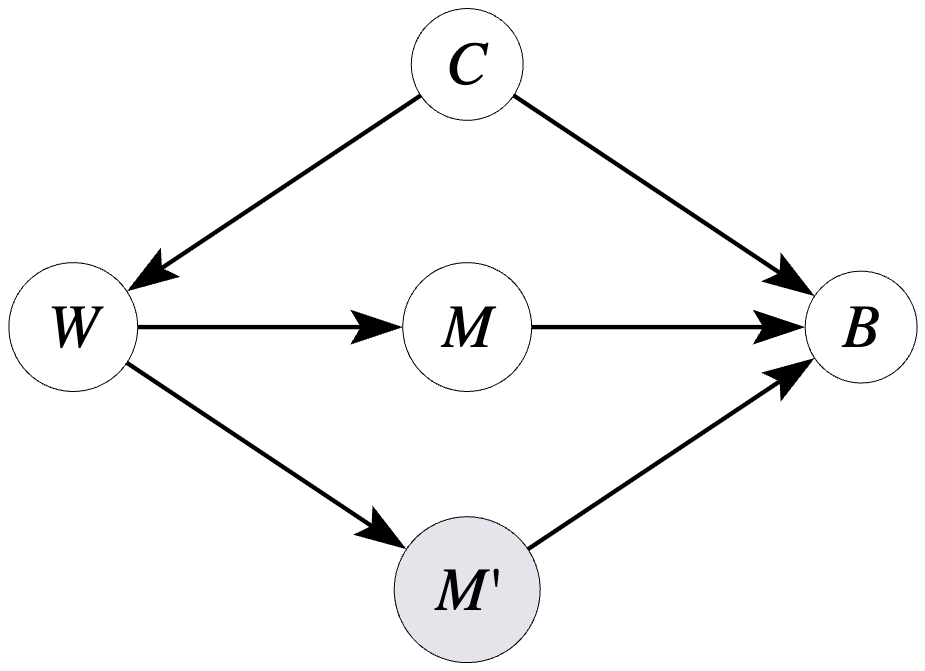}
\caption{Partially observed mediators.}
\label{fig:PO-mediator}
\end{wrapfigure}

\Cref{fig:spikes-decomposition} shows this decomposition. In each experimental setting we may observe the above spikes, possibly in partial. Below, we consider different observability scenarios, and discuss epiphenomenality in each instance. In each example, the set $\mathbf{S}$ is the observed spikes and $\mathbf{U}$ contains all other factors, including the unobserved spikes.

\textbf{Partially observed mediators.} Suppose $\mathbf{M} \not\subset \mathbf{S}$, meaning that there exists some mediators that are not observed in $\mathbf{S}$. Assuming all other factors are observed, we describe this situation with the graph in \Cref{fig:PO-mediator}. There exists direct causal paths $ \mathbf{W} \to M \to \mathbf{B}$ for every $M \in \mathbf{M}\setminus \mathbf{S}$. Therefore, interventions on $\mathbf{W}$ affect $M$ and the changes in $M$ carries over to $\mathbf{B}$, which implies that waves are causal to behavior in this case. This intuition is justified below.

\begin{proposition} \label{prop:directed-path} [From causal paths to causal waves] Consider an SCM over the sets of variables $\mathbf{W},\mathbf{B},\mathbf{S},\mathbf{U}$. If there exists a directed path from a wave $W \in \mathbf{W}$ to a behavior $B \in \mathbf{B}$, such as,
\begin{equation*}
    W \to U_1 \to U_2 \to  ...\to U_n \to B,
\end{equation*}
where $U_1, U_2, \dots U_n \in \mathbf{U}$, then,
\begin{equation}
    P(\mathbf{B} \mid \mathbf{S}; do(\mathbf{W})) \neq P(\mathbf{B} \mid \mathbf{S}).
\end{equation}
\end{proposition}

In words, waves exert a causal effect on behavior whenever not all directed paths from waves to behavior are blocked by observation of at least one variable along each path (i.e., included in $\mathbf{S}$). Below we formally define this notion as causal relevance of a scientific phenomenon.

\begin{definition} [Epiphenomenality] A set of variables $\mathbf{Z}$ is epiphenomenal to the outcome $Y$ given the predictors $\mathbf{X}$ if,
\begin{equation}
    P(\mathbf{Y} \mid \mathbf{X}; do(\mathbf{Z})) = P(\mathbf{Y}\mid \mathbf{X}),
\end{equation}
i.e., the passive distribution of outcome $\mathbf{Y}$ conditional on the value of predictors $\mathbf{X}$ remains invariant under interventions on the variables $\mathbf{Z}$.
\end{definition}

\begin{wrapfigure}{r}{0.2\textwidth}
\centering
\includegraphics[width=\linewidth]{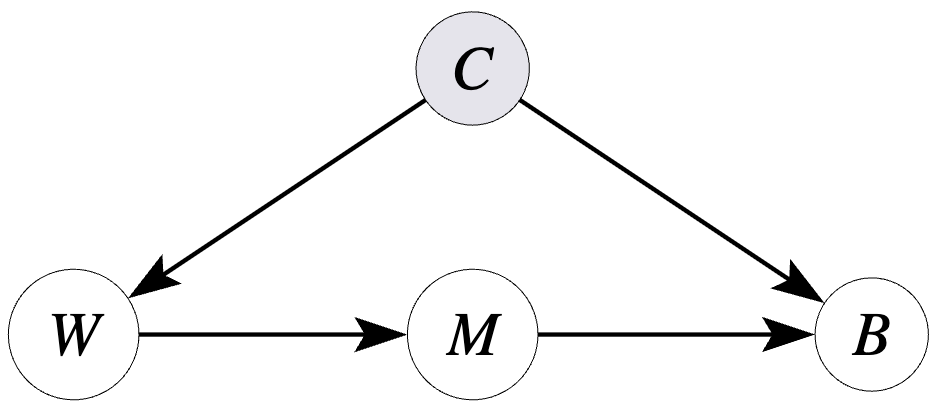}
\caption{Front-door graph}
\label{fig:frontdoor}
\end{wrapfigure}

\textbf{Front-door graph.} Suppose all mediators are observed (i.e., $\mathbf{M} \subset \mathbf{S}$), but $\mathbf{C} \subset \mathbf{U}$, meaning that none of the common causes of waves and behavior are observed. The graph in \Cref{fig:frontdoor} describes this situation. One might claim that since all of the causal influence of waves on behavior is mediated by observed spikes, it deems waves epiphenomenal. This interpretation is not accurate, as stated formally below.

\begin{proposition} \label{prop:front-door} [Causal waves in front-door graph] If the graph in \Cref{fig:frontdoor} is induced by the true SCM that governs the brain activities, then the waves would have causal influence on behavior. Formally speaking,
\begin{equation}
    P(\mathbf{B} \mid \mathbf{S}; do(\mathbf{W})) \neq P(\mathbf{B} \mid \mathbf{S})
\end{equation}
holds for all SCMs that induces the front-door graph, except for a measure zero subset.

\textbf{\textit{proof sketch.}} \normalfont Using do-calculus we can not derive the r.h.s. from the l.h.s., and since do-calculus is a complete derivation system (\cite{huang2012pearl}), the above holds for all nondegenerate SCMs. 
\end{proposition}

\Cref{prop:directed-path,prop:front-door} apply to only specific graphs (or class of graphs), while the next result is a general graphical characterization of epiphenomenality which applies to any causal graph that we have not considered here.

\begin{theorem} \label{thm:epiph-via-rule3} Let $Z,X,Y$ be disjoint sets of variables in a causal diagram $\mathcal{G}$. A set of variables $Z$ is epiphenomenal to $Y$ given $X$ for all SCMs that induce the graph $\mathcal{G}$, if,
\begin{equation}
    Z \indep_{d} Y \mid X \text{ in } \mathcal{G}_{\overline
    {Z(X)}},
\end{equation}
where $Z(X)$ are nodes in $Z$ that are not ancestor to any nodes in $X$, and $\mathcal{G}_{\overline
    {Z(X)}}$ denotes the graph obtained from removing the arrows that point to the set of variables $Z(X)$.
\end{theorem}

\begin{wrapfigure}{r}{0.2\textwidth}
\centering
\includegraphics[width=\linewidth]{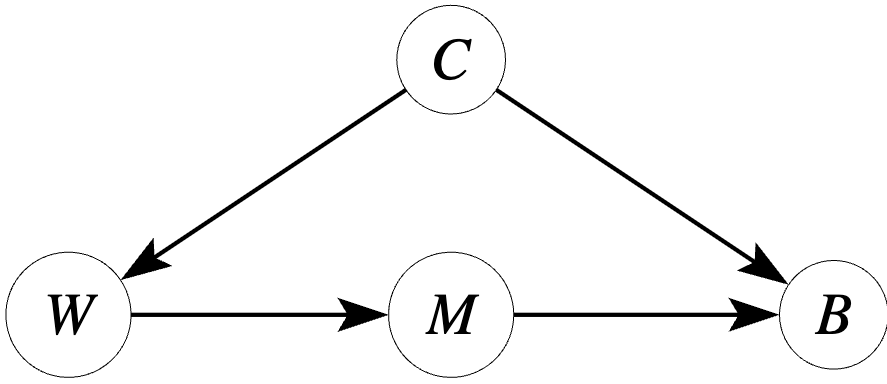}
\caption{Epiphenomenality graph.}
\label{fig:all-obs}
\end{wrapfigure}

Attested by \Cref{thm:epiph-via-rule3}, a certain case of epiphenomenality is when all confounders and mediators from \Cref{fig:spikes-decomposition} are observed (\Cref{fig:all-obs}), which implies $P(\mathbf{B}\mid \mathbf{C},\mathbf{M}; do(\mathbf{W})) = P(\mathbf{B}\mid \mathbf{C},\mathbf{M})$.

\section*{Discussion}


This paper reframes the spike--wave debate in causal terms. Rather than asking whether waves matter more or less than spikes, we show how to decide when any signal matters at all. We define epiphenomenality as invariance of interventional distributions, provide a graph-based certificate for collapsing variables without causal loss, and demonstrate that the causal role of waves depends on the assumed graph.
The key message is twofold. First, decoding success does not establish mechanism: prediction and explanation live at different levels. Second, causal claims require explicit diagrams and identification targets. Our framework clarifies when waves can be treated as summaries of spikes and when such abstraction erases mechanism.
Practically, this shifts multiscale neuroscience from post-hoc debates to explicit causal programs: state assumptions, design experiments to test $P(B;\mathrm{do}(W))$ versus $P(B)$, and report results relative to the graph. 
While our formulation treats hidden causes abstractly within the exogenous space of the structural model, future theoretical extensions could expand this treatment by formalizing the role of latent confounders more explicitly. Approaches such as Instrumental Variable models or latent factor frameworks \cite{wang2019blessings, wang2020towards} could, in principle, be embedded within our causal framing to represent structured hidden variables rather than unmodeled noise. Such extensions would not alter the conceptual component of our argument but would refine how unobserved causal structure is represented, bridging our formalism with probabilistic models of latent neural dynamics. Challenges such as hidden confounders, recursive coupling, and nonstationarity persist, yet within this framework they appear not as ambiguities but as declared assumptions, making the boundaries of inference explicit rather than contested.

In closing, the goal is not to decide whether waves cause behavior. It is to supply the tools for asking that question rigorously. By grounding the debate in causal model, interventions, and graphs, we return to the abstract's claim: the value of a signal lies not in its correlates, but in the causal answers it makes possible.

\newpage
{
\bibliographystyle{unsrtnat}
\bibliography{CameraReady}
}
\newpage
\appendix

\section{Background}

This section provides the necessary background on causal inference concepts that underpin our formal analysis of spike-wave duality. We introduce Structural Causal Models (SCMs), causal graphs, interventional distributions, and do-calculus as the foundational tools for reasoning about causality in neuroscientific contexts.

\textbf{Notation:} Throughout this paper, we use bold capital letters (e.g., $\mathbf{W}, \mathbf{S}, \mathbf{B}$) to denote sets of variables, regular capital letters (e.g., $W, S, B$) for individual variables, and lowercase letters (e.g., $w, s, b$) for specific values of variables. For notational simplicity in the main text, we write $do(\mathbf{W})$ as shorthand for $do(\mathbf{W} \gets \mathbf{w})$ when the specific intervention values are clear from context.

\subsection{Structural Causal Models}

A Structural Causal Model (SCM) provides a mathematical framework for representing causal relationships between variables \cite{Pearl2009}. An SCM is defined as a tuple $\mathcal{M} = \langle \mathbf{U}, \mathbf{V}, \mathbf{F}, P(\mathbf{U}) \rangle$ where:

\begin{itemize}
    \item $\mathbf{U}$ is a set of exogenous (unobserved) variables that are determined by factors outside the model
    \item $\mathbf{V} = \{V_1, V_2, \ldots, V_n\}$ is a set of endogenous (observed) variables that are determined by variables in the model
    \item $\mathbf{F} = \{f_1, f_2, \ldots, f_n\}$ is a set of functions, where each $f_i$ determines the value of $V_i$ in terms of its parents: $V_i \gets f_i(PA_i, U_i)$, where $PA_i \subset \mathbf{V} \setminus \{V_i\}$ are the parents of $V_i$ and $U_i \subset \mathbf{U}$
    \item $P(\mathbf{U})$ is a probability distribution over the exogenous variables
\end{itemize}

The structural equations encode the causal mechanisms of the system. For instance, in the context of neural activity, we might have:
\begin{align}
    W &\gets f_W(\mathbf{S}_{\text{upstream}}, U_W) \\
    S &\gets f_S(W, \mathbf{S}_{\text{other}}, U_S) \\
    B &\gets f_B(\mathbf{S}, W, U_B)
\end{align}
where $W$ represents a wave variable, $S$ represents a spike variable, $B$ represents behavior, and $U_W, U_S, U_B$ are exogenous noise terms.

\subsection{Causal Graphs and d-separation}

Every SCM induces a directed acyclic graph (DAG) $\mathcal{G} = (\mathbf{V}, \mathbf{E})$ where vertices correspond to endogenous variables and directed edges represent direct causal relationships. An edge $V_i \to V_j$ exists if and only if $V_i \in PA_j$.

The concept of d-separation (directional separation) provides a graphical criterion for determining conditional independence relationships encoded by the causal graph \cite{Pearl1988}.

\begin{definition}[d-separation]
A path $p$ between nodes $X$ and $Y$ in a DAG $\mathcal{G}$ is blocked by a set of nodes $\mathbf{Z}$ if and only if:
\begin{enumerate}
    \item $p$ contains a chain $i \to m \to j$ or a fork $i \leftarrow m \to j$ such that the middle node $m$ is in $\mathbf{Z}$, or
    \item $p$ contains a collider $i \to m \leftarrow j$ such that the collision node $m$ is not in $\mathbf{Z}$ and no descendant of $m$ is in $\mathbf{Z}$.
\end{enumerate}
If all paths between $X$ and $Y$ are blocked by $\mathbf{Z}$, then $X$ and $Y$ are d-separated by $\mathbf{Z}$, denoted $X \indep_d Y \mid \mathbf{Z}$.
\end{definition}

The fundamental connection between d-separation and probabilistic independence is:

\begin{theorem}[Global Markov Property]
If $\mathcal{G}$ is the causal graph induced by SCM $\mathcal{M}$, then for any disjoint sets of variables $\mathbf{X}, \mathbf{Y}, \mathbf{Z}$:
\begin{equation}
    \mathbf{X} \indep_d \mathbf{Y} \mid \mathbf{Z} \text{ in } \mathcal{G} \Rightarrow \mathbf{X} \indep \mathbf{Y} \mid \mathbf{Z} \text{ in } P
\end{equation}
where $P$ is the probability distribution induced by $\mathcal{M}$.
\end{theorem}

\subsection{Interventions and the $do(x)$ operator}

The key innovation of the causal inference framework is the formal treatment of interventions. An intervention on a variable $X$ involves setting its value externally, overriding the natural causal mechanism that would normally determine $X$.

\begin{definition}[Intervention and do-operator]
An intervention $do(X \gets x)$ on variable $X$ in SCM $\mathcal{M}$ produces a new model $\mathcal{M}_x$ where:
\begin{itemize}
    \item The structural equation for $X$ is replaced by $X \gets x$ (a constant)
    \item All other structural equations remain unchanged
    \item The corresponding graph $\mathcal{G}_x$ is obtained by removing all incoming edges to $X$
\end{itemize}
The interventional distribution $P(Y ; do(X \gets x))$ is the distribution of $Y$ under this modified model.
\end{definition}

For our application, $P(\mathbf{B} ; do(\mathbf{W} \gets \mathbf{w}))$ represents the distribution of behavior when waves are experimentally set to specific values $\mathbf{w}$, breaking any natural dependencies that would normally determine wave activity.

The fundamental distinction is that generally:
\begin{equation}
    P(Y \mid X = x) \neq P(Y ; do(X \gets x))
\end{equation}
The left side represents seeing $X = x$ (passive observation), while the right side represents making $X \gets x$ (active intervention).

\subsection{do-calculus}

do-calculus provides three rules for deriving interventional distributions from observational data, given a causal graph \cite{Pearl1995, Shpitser2006}:

\textbf{Rule 1 (Insertion/deletion of observations):}
\begin{equation}
    P(\mathbf{Y} ; do(\mathbf{X} \gets \mathbf{x}), \mathbf{Z} = \mathbf{z}, \mathbf{W} = \mathbf{w}) = P(\mathbf{Y} ; do(\mathbf{X} \gets \mathbf{x}), \mathbf{W} = \mathbf{w})
\end{equation}
if $(\mathbf{Y} \indep_d \mathbf{Z} \mid \mathbf{X}, \mathbf{W})_{\mathcal{G}_{\overline{\mathbf{X}}}}$

\textbf{Rule 2 (Action/observation exchange):}
\begin{equation}
    P(\mathbf{Y} ; do(\mathbf{X} \gets \mathbf{x}), do(\mathbf{Z} \gets \mathbf{z}), \mathbf{W} = \mathbf{w}) = P(\mathbf{Y} ; do(\mathbf{X} \gets \mathbf{x}), \mathbf{Z} =\mathbf{z}, \mathbf{W} = \mathbf{w})
\end{equation}
if $(\mathbf{Y} \indep_d \mathbf{Z} \mid \mathbf{X}, \mathbf{W})_{\mathcal{G}_{\overline{\mathbf{X}}, \underline{\mathbf{Z}}}}$

\textbf{Rule 3 (Insertion/deletion of actions):}
\begin{equation}
    P(\mathbf{Y} ; do(\mathbf{X} \gets \mathbf{x}), do(\mathbf{Z} \gets \mathbf{z}), \mathbf{W} \gets \mathbf{w}) = P(\mathbf{Y} ; do(\mathbf{X} \gets \mathbf{x}), \mathbf{W} = \mathbf{w})
\end{equation}
if $(\mathbf{Y} \indep_d \mathbf{Z} \mid \mathbf{X}, \mathbf{W})_{\mathcal{G}_{\overline{\mathbf{X}}, \overline{\mathbf{Z}(\mathbf{W})}}}$

where $\mathcal{G}_{\overline{\mathbf{X}}}$ removes all incoming edges to $\mathbf{X}$, $\mathcal{G}_{\underline{\mathbf{Z}}}$ removes all outgoing edges from $\mathbf{Z}$, and $\mathbf{Z}(\mathbf{W})$ are nodes in $\mathbf{Z}$ that are not ancestors of any node in $\mathbf{W}$.

These rules are complete: if a causal query can be computed from observational data given a graph, do-calculus will find the derivation.

\subsection{The Causal Hierarchy Theorem}

The Causal Hierarchy Theorem establishes fundamental limitations on what can be learned from observational data alone \cite{bareinboim-r60}.

\begin{theorem}[Causal Hierarchy Theorem]
Causal inference problems form a hierarchy:
\begin{enumerate}
    \item \textbf{Association} ($P(Y \mid X = x)$): Statistical dependence observable in passive data
    \item \textbf{Intervention} ($P(Y ; do(X \gets x))$): Effects of deliberate actions
    \item \textbf{Counterfactuals} ($P(Y_x \mid X = x', Y = y')$): Retrospective reasoning about alternatives
\end{enumerate}
For almost all SCMs, even with infinite samples from quantities at the level $i$, one can not be uniquely determine the quantities at level $j > i$.
\end{theorem}

This theorem directly applies to the spike-wave debate: no amount of observational correlation data can definitively establish whether waves are epiphenomenal without additional causal assumptions or interventional experiments.


\section{Extended Discussion}

\subsection{Theoretical Implications}

Our causal framework for spike-wave duality represents a methodological shift from correlation-based analysis to principled causal inference. Rather than taking sides in the longstanding debate about whether neural waves are causally relevant or epiphenomenal, we provide formal tools to adjudicate between competing hypotheses. The importance of explicitly stating these competing hypotheses cannot be overstated—only by formalizing different causal structures can we design experiments capable of distinguishing between them. This approach moves neuroscience away from premature commitment to either position without proper causal analysis, toward a more rigorous framework for understanding multi-scale neural phenomena.

\subsection{Challenges in Causal Discovery of Spike-Wave Interactions}

While our theoretical framework provides a principled foundation, practical implementation in neural data analysis faces several fundamental challenges that must be addressed for successful application.

\subsubsection{Unavoidable Hidden Nodes}

The contribution of individual neurons to behavior is often limited, and proper causal graph representation would require millions of neurons as nodes. In practice, neural data contains numerous confounding variables that cannot be directly measured or controlled. These include neurons falling outside the recording window, neurons removed due to noise issues, and broader network effects that influence observed signals. Such unobserved confounders can significantly influence the observed neural signals, making it challenging to establish direct causal relationships \cite{Pearl2009}.

Potential solutions include studying causal effects in more controlled environments where confounding factors can be better identified, such as organoids or sliced brain tissue preparations. For in-vivo datasets, methods that explicitly account for existing confounding factors, such as those developed by Gerhardus et al. \cite{pfadenhauer2017making}, would be beneficial. However, the fundamental challenge of hidden confounders remains a significant limitation for causal discovery in complex neural systems.

\subsubsection{Cyclic Causal Graphs: Recursive Spike-Wave Interactions}

Biophysical studies and expert knowledge about spike-field potential interactions at the synaptic level indicate that these interactions are inherently recursive. Spikes influence local field potentials, which in turn can influence subsequent spiking activity through various mechanisms including ephaptic coupling and network-level feedback. This creates cyclic causal relationships that are mathematically much more challenging to analyze than acyclic graphs \cite{spirtes2000}.

Several approaches can address this challenge. Existing causal discovery methods like PCMCI (Peter and Clark Momentary Conditional Independence) are specifically designed to handle cyclic graphs \cite{runge2019}. Alternatively, one can decouple spiking activity from peripheral field potentials by isolating specific components of neural activity for independent analysis, though this approach may sacrifice some biological realism for analytical tractability.

\subsubsection{Mixed Linear and Non-Linear Interactions}

Neural signals exhibit both linear and non-linear interactions, which are critical for understanding brain function \cite{mcintosh_nonlin, Dayan2001}. In the context of spike-wave interactions, the dynamics may be particularly complex: brain wave interactions across regions might be modeled as relatively straightforward linear interactions, whereas spike interactions across regions are often inherently non-linear \cite{izhikevich_synchronization, hyafil_coupling}.

The theoretical relationship between presynaptic potentials and spiking activity is fundamentally non-linear due to the sigmoid nature of neuronal activation functions \cite{beuter_nonlin, markram_plasticity}. This complexity poses significant challenges for causal discovery methods. Many traditional algorithms, such as Granger causality, are designed for linear systems and fail to capture the full complexity of neural dynamics \cite{glymour_causal, pearl_causal, bressler_seth}. Future applications of our framework will require causal discovery methods that can accommodate both linear and non-linear dynamics simultaneously.

\subsubsection{High Autocorrelation of Neural Signals}

Autocorrelation—the correlation of a signal with delayed copies of itself—is ubiquitous in neural signals due to intrinsic properties of neuronal firing patterns and synaptic connections \cite{mcintosh2008}. This temporal dependence violates the independence assumptions underlying many causal discovery algorithms, potentially leading to spurious causal inferences.

Addressing this challenge requires time-series analysis methods that explicitly account for autocorrelation. Approaches include autoregressive integrated moving average (ARIMA) models and other methods designed for handling serial dependencies in data. However, incorporating these methods into causal discovery frameworks while maintaining the ability to distinguish genuine causal relationships from autocorrelation-induced spurious associations remains an active area of research.

\subsubsection{Violation of Stationarity Assumptions Across Cognitive States}

Stationarity assumes that the statistical properties of a signal remain constant over time. Neural signals frequently violate this assumption across different cognitive states, with properties such as mean, variance, and spectral characteristics changing substantially \cite{breakspear2005}. This non-stationarity can confound causal discovery by making it difficult to distinguish genuine causal relationships from state-dependent correlations.

Simple solutions to uncontrolled nonstationarity often include restricting analysis to intervals with similar cognitive states, though this approach may limit the generalizability of findings. More sophisticated approaches involve non-stationary signal processing techniques such as adaptive filtering or time-varying auto-regressive models. The heterogeneity can in fact aid causal discovery as well; any perturbation to the causal mechanisms reveals information about the structure beyond what is identifiable from mere observations, even if the perturbations are imperfect and due to the nature or unknown sources. By breaking the observational symmetry between the cause and the effect, some existing methods can recover the causal graph beyond what is achievable with perfectly stationary data \cite{perry2022causal, peters2016causal, jaber2020causal, ghassami2017learning, mooij2020joint, huang2020causal,li2023discovery,jalaldoust2025multidomain,peters2016causal}, but our current formulation does not capture such nuanced cases involving multi-domain data.

\subsubsection{Spatial Sampling Scale}

The historical skepticism toward oscillations stems largely from the coarse spatial sampling of field potential recordings. Conventional electrodes aggregate signals from thousands of neurons, blurring fine-scale structure and making waves appear as diffuse population averages rather than mechanistically grounded dynamics. This resolution gap fostered the view that oscillations cannot exert precise computational influence on single neurons, casting them as epiphenomenal reflections of spiking rather than causal contributors to it.

The fabrication of high-density, high-resolution recording arrays now directly addresses one of the central limitations outlined in the introduction—the coarse spatial sampling that historically relegated oscillations to the status of low-resolution, and thus supposedly epiphenomenal, signals. Platforms such as the Neuropixels probes \cite{jun2017fully} and the BISC microelectrode arrays \cite{jung2025stable} allow simultaneous recording of spikes and field potentials across contiguous cortical regions with cellular precision. This convergence eliminates the scale mismatch that once separated the two descriptions, enabling oscillations to be studied as spatially structured, mechanistically grounded phenomena rather than population averages. As these technologies mature, they may reveal whether the apparent coarse structure of waves was a limitation of measurement rather than a property of the brain itself—transforming the spike–wave debate from one constrained by resolution to one grounded in causal interpretation.

\subsection{Future Directions}

Despite these challenges, our framework opens several promising avenues for future research. First, developing computational tools that implement our theoretical contributions for real neural data, with particular attention to the challenges outlined above. Second, designing controlled experiments that can test specific predictions about when waves are epiphenomenal versus causally sufficient. Third, extending the approach to other multi-scale phenomena in neuroscience beyond spike-wave duality. Finally, investigating how our causal abstraction principles might apply to other complex systems where multiple levels of description compete for explanatory primacy.

The ultimate goal is not to definitively resolve the spike-wave debate, but to provide neuroscience with principled tools for making such determinations based on rigorous causal analysis rather than correlational evidence alone.

\section{Proofs}

\subsection{Proof of Proposition \ref{prop:directed-path}}

\begin{proof}
Consider a directed path $\pi: W \to U_1 \to U_2 \to \cdots \to U_n \to B$ where $W \in \mathbf{W}$, $B \in \mathbf{B}$, and $U_i \in \mathbf{U}$ for all $i \in \{1, \ldots, n\}$. For any variable $V$ in a causal graph, we denote by $PA_V$ the parent set of $V$, defined as the set of all variables that have a direct causal edge pointing into $V$. In the context of an SCM, these parents appear as arguments in the structural equation that determines $V$.

The structural equations along the path take the following form. For the first unobserved variable, we have $U_1 \gets f_{U_1}(W, PA_{U_1} \setminus \{W\}, \epsilon_{U_1})$, where $PA_{U_1} \setminus \{W\}$ represents all parents of $U_1$ except $W$, and $\epsilon_{U_1}$ is an exogenous noise term drawn from the background distribution. For intermediate variables along the path, $U_i \gets f_{U_i}(U_{i-1}, PA_{U_i} \setminus \{U_{i-1}\}, \epsilon_{U_i})$ for $i \in \{2, \ldots, n\}$. Finally, the behavior variable satisfies $B \gets f_B(U_n, PA_B \setminus \{U_n\}, \epsilon_B)$, where again we explicitly separate the contribution from the path variable $U_n$ from other potential parents.

When we perform the intervention $do(W \gets w)$, we fundamentally alter the data-generating process. The structural equation for $W$ is replaced entirely by the constant assignment $W \gets w$, removing all edges into $W$ in the causal graph. This is not mere conditioning—it is a surgical modification of the causal structure. Under this intervention, the distribution of $U_1$ becomes
\begin{equation}
P(U_1 = u_1 ; do(W \gets w)) = \int P(\epsilon_{U_1} = e) \cdot 1[f_{U_1}(w, PA_{U_1} \setminus \{W\}, e) = u_1] \, de
\end{equation}
where $1[\cdot]$ is the indicator function. This distribution differs from the natural distribution $P(U_1)$ whenever the partial derivative $\partial f_{U_1}/\partial W$ is non-zero, which holds for all non-degenerate SCMs where $W$ actually influences $U_1$.

The change in $U_1$'s distribution propagates forward through the path. Since $U_2$ depends on $U_1$ through its structural equation, and the intervention has altered $P(U_1)$, we obtain a modified distribution $P(U_2 ; do(W \gets w))$ that differs from $P(U_2)$. This cascade continues: each $U_i$ inherits the perturbation from its predecessor $U_{i-1}$, ultimately reaching $B$ through its dependence on $U_n$.

Now we consider what happens when we condition on the observed variables $\mathbf{S}$. The critical insight is that conditioning cannot block a causal path unless we condition on at least one variable along that path. Since all intermediate variables $U_1, \ldots, U_n$ belong to the unobserved set $\mathbf{U}$, they are by definition not contained in $\mathbf{S}$. The causal influence from $W$ to $B$ flows through these unobserved intermediaries unimpeded.

To formalize this mathematically, we decompose the conditional distribution using the law of total probability:
\begin{equation}
P(B = b \mid \mathbf{S} = \mathbf{s}; do(W \gets w)) = \int P(B = b \mid \mathbf{S} = \mathbf{s}, U_n = u) \cdot P(U_n = u \mid \mathbf{S} = \mathbf{s}; do(W \gets w)) \, du
\end{equation}

The first term $P(B = b \mid \mathbf{S} = \mathbf{s}, U_n = u)$ is determined by the structural equation for $B$ and does not depend on whether $W$ was intervened upon or merely observed. However, the second term $P(U_n = u \mid \mathbf{S} = \mathbf{s}; do(W \gets w))$ explicitly depends on the intervention. Since the path from $W$ to $U_n$ passes only through unobserved variables, this distribution differs from the observational $P(U_n = u \mid \mathbf{S} = \mathbf{s})$.

Specifically, in the observational case, $U_n$ 's distribution reflects the natural variation in $W$ and all intermediate variables. Under intervention, $W$ is held fixed at $w$, and this constraint propagates through to alter $U_n$'s distribution. For any SCM where the composed function from $W$ to $U_n$ has non-zero derivative—that is, where the causal effect actually exists—these two distributions are distinct. Therefore, $P(\mathbf{B} \mid \mathbf{S}; do(\mathbf{W})) \neq P(\mathbf{B} \mid \mathbf{S})$.
\end{proof}

\subsection{Proof of Proposition \ref{prop:front-door}}

\begin{proof}
The front-door graph encodes a specific pattern of causal relationships: waves influence behavior exclusively through observed mediators $\mathbf{M} \subseteq \mathbf{S}$, while unobserved confounders $\mathbf{C} \subseteq \mathbf{U}$ create a backdoor path between waves and behavior. Formally, the graph contains edges $\mathbf{W} \to \mathbf{M}$, $\mathbf{M} \to \mathbf{B}$, $\mathbf{C} \to \mathbf{W}$, and $\mathbf{C} \to \mathbf{B}$, with no direct edge from $\mathbf{W}$ to $\mathbf{B}$.

We seek to compare the interventional distribution $P(\mathbf{B} \mid \mathbf{S}; do(\mathbf{W}))$ with the observational distribution $P(\mathbf{B} \mid \mathbf{S})$. For clarity, we consider the pure front-door case where $\mathbf{S} = \mathbf{M}$. We begin by deriving the interventional distribution through a systematic application of do-calculus.

Starting with $P(\mathbf{B} \mid \mathbf{S}; do(\mathbf{W}))$, we systematically apply the rules of do-calculus to derive an expression in terms of observational quantities. We first apply Rule 2, which allows us to exchange action and observation when appropriate d-separation conditions hold. Since $(\mathbf{S} \perp \mathbf{B} \mid \mathbf{W})$ in $\mathcal{G}_{\overline{\mathbf{W}\mathbf{S}}}$ (the graph with incoming edges to both $\mathbf{W}$ and $\mathbf{S}$ removed), we obtain:
\begin{equation}
P(\mathbf{B} \mid \mathbf{S}; do(\mathbf{W})) = P(\mathbf{B}; do(\mathbf{W}), do(\mathbf{S}))
\end{equation}

Applying Rule 3 to remove the intervention on $\mathbf{W}$, we use the fact that $(\mathbf{W} \perp \mathbf{B} \mid \mathbf{S})$ in $\mathcal{G}_{\overline{\mathbf{S}\mathbf{W}}}$. Since all paths from $\mathbf{W}$ to $\mathbf{B}$ pass through $\mathbf{S} = \mathbf{M}$ after removing incoming edges, we have:
\begin{equation}
P(\mathbf{B}; do(\mathbf{W}), do(\mathbf{S})) = P(\mathbf{B}; do(\mathbf{S}))
\end{equation}

We now express this marginal distribution by summing over all possible values of $\mathbf{W}$:
\begin{equation}
P(\mathbf{B}; do(\mathbf{S})) = \sum_{\mathbf{w}'} P(\mathbf{B}, \mathbf{W} = \mathbf{w}'; do(\mathbf{S}))
\end{equation}

Applying the chain rule to decompose the joint distribution:
\begin{equation}
P(\mathbf{B}, \mathbf{W} = \mathbf{w}'; do(\mathbf{S})) = P(\mathbf{B} \mid \mathbf{W} = \mathbf{w}'; do(\mathbf{S})) \cdot P(\mathbf{W} = \mathbf{w}'; do(\mathbf{S}))
\end{equation}

For the conditional term, we apply Rule 3. Since $(\mathbf{S} \perp \mathbf{W})$ in $\mathcal{G}_{\overline{\mathbf{S}}}$ (after removing incoming edges to $\mathbf{S}$, there are no common ancestors), we get:
\begin{equation}
P(\mathbf{B} \mid \mathbf{W} = \mathbf{w}'; do(\mathbf{S})) = P(\mathbf{B} \mid \mathbf{S}, \mathbf{W} = \mathbf{w}')
\end{equation}

Similarly, for the marginal term:
\begin{equation}
P(\mathbf{W} = \mathbf{w}'; do(\mathbf{S})) = P(\mathbf{W} = \mathbf{w}')
\end{equation}

Finally, applying Rule 2 to convert the intervention on $\mathbf{S}$ back to conditioning (using $(\mathbf{S} \perp \mathbf{B} \mid \mathbf{W})$ in $\mathcal{G}_{\overline{\mathbf{S}}}$), we arrive at:
\begin{equation}
P(\mathbf{B} \mid \mathbf{S}; do(\mathbf{W})) = \sum_{\mathbf{w}'} P(\mathbf{B} \mid \mathbf{S}, \mathbf{W} = \mathbf{w}') P(\mathbf{W} = \mathbf{w}')
\end{equation}

The interventional distribution is a weighted average of the observational conditional $P(\mathbf{B} \mid \mathbf{S}, \mathbf{W} =\mathbf{w}')$ over all possible values $\mathbf{w}'$, weighted by the marginal distribution $P(\mathbf{W} = \mathbf{w}')$.
For epiphenomenality, the following must hold,
\begin{equation}
P(\mathbf{B} \mid \mathbf{S}) \overset{?}{=} \sum_{\mathbf{w}'} P(\mathbf{B} \mid \mathbf{S}, \mathbf{W} = \mathbf{w}') P(\mathbf{W} = \mathbf{w}').
\end{equation}

The only scenarios where this equality could hold observationally are degenerate cases: either the confounding effects are exactly zero (removing the backdoor path entirely), or multiple effects cancel through a precise balance of parameters that has measure zero in the space of all SCMs. For generic SCMs where confounding effects are non-zero and non-canceling, we conclude that $P(\mathbf{B} \mid \mathbf{S}; do(\mathbf{W})) \neq P(\mathbf{B} = \mathbf{S})$.
\end{proof}

\subsection{Proof of Theorem \ref{thm:epiph-via-rule3}}

\begin{proof}
We prove that variables $\mathbf{Z}$ are epiphenomenal to outcomes $\mathbf{Y}$ given predictors $\mathbf{X}$—formally, that $P(\mathbf{Y} \mid \mathbf{X}; do(\mathbf{Z})) = P(\mathbf{Y} \mid \mathbf{X})$—if and only if $\mathbf{Z} \indep_d \mathbf{Y} \mid \mathbf{X}$ in the graph $\mathcal{G}_{\overline{\mathbf{Z}(\mathbf{X})}}$.

Before proceeding, we must carefully define our terms. Let $\mathbf{Z}(\mathbf{X})$ denote the subset of $\mathbf{Z}$ consisting of variables that are not ancestors of any variable in $\mathbf{X}$. Formally, $Z \in \mathbf{Z}(\mathbf{X})$ if and only if there exists no directed path from $Z$ to any $X \in \mathbf{X}$ in the original causal graph $\mathcal{G}$. The modified graph $\mathcal{G}_{\overline{\mathbf{Z}(\mathbf{X})}}$ is constructed from $\mathcal{G}$ by removing all incoming edges to variables in $\mathbf{Z}(\mathbf{X})$. This construction represents the post-intervention graph for a specific subset of $\mathbf{Z}$—those that cannot influence the conditioning set $\mathbf{X}$.

We first establish the forward direction. Assume that $\mathbf{Z} \indep_d \mathbf{Y} \mid \mathbf{X}$ holds in $\mathcal{G}_{\overline{\mathbf{Z}(\mathbf{X})}}$. This d-separation statement means that in the modified graph where we have removed incoming edges to non-ancestors of $\mathbf{X}$ in $\mathbf{Z}$, all paths from $\mathbf{Z}$ to $\mathbf{Y}$ are blocked by $\mathbf{X}$. 

Rule 3 of do-calculus provides conditions for deleting interventions from conditional distributions. Specifically, it states that $P(\mathbf{Y} \mid \mathbf{X}; do(\mathbf{Z}), \mathbf{W}) = P(\mathbf{Y} \mid \mathbf{X}, \mathbf{W})$ if $\mathbf{Y} \indep_d \mathbf{Z} \mid \mathbf{X}, \mathbf{W}$ in $\mathcal{G}_{\overline{\mathbf{X}}, \overline{\mathbf{Z}(\mathbf{W})}}$. In our case, with empty $\mathbf{W}$, the condition simplifies to requiring $\mathbf{Y} \indep_d \mathbf{Z} \mid \mathbf{X}$ in $\mathcal{G}_{\overline{\mathbf{Z}(\mathbf{X})}}$, which is precisely our assumption. Therefore, Rule 3 allows us to conclude:
\begin{equation}
P(\mathbf{Y} \mid \mathbf{X}; do(\mathbf{Z})) = P(\mathbf{Y} \mid \mathbf{X})
\end{equation}
establishing that $\mathbf{Z}$ is epiphenomenal to $\mathbf{Y}$ given $\mathbf{X}$.

For the reverse direction, suppose that $P(\mathbf{Y} \mid \mathbf{X}; do(\mathbf{Z})) = P(\mathbf{Y} \mid \mathbf{X})$ holds for all SCMs compatible with causal graph $\mathcal{G}$. This universal validity is crucial—the equality must hold not just for some particular SCM, but for every SCM that could generate the graph structure $\mathcal{G}$. 

The completeness theorem for do-calculus, established by Huang and Valtorta (2006) and independently by Shpitser and Pearl (2006), states that if a causal effect is identifiable from a graph, then it can be computed using the three rules of do-calculus. More precisely, if an equality between interventional distributions holds for all SCMs compatible with a graph, then this equality must be derivable using do-calculus rules.

In our case, the equality $P(\mathbf{Y} \mid \mathbf{X}; do(\mathbf{Z})) = P(\mathbf{Y} \mid \mathbf{X})$ represents the deletion of an intervention operator. Among the three rules of do-calculus, only Rule 3 permits such deletion. Rules 1 and 2 deal with insertion/deletion of observations and action/observation exchange, respectively, but cannot eliminate a $do(\cdot)$ operator entirely from a distribution.

Since the equality must be derivable via do-calculus, and only Rule 3 can derive it, the graphical condition required by Rule 3 must hold. This condition is precisely $\mathbf{Y} \indep_d \mathbf{Z} \mid \mathbf{X}$ in $\mathcal{G}_{\overline{\mathbf{Z}(\mathbf{X})}}$, completing our proof of the reverse direction.

The theorem provides a complete graphical characterization of epiphenomenality. Variables $\mathbf{Z}$ are epiphenomenal to $\mathbf{Y}$ given $\mathbf{X}$ when their influence on $\mathbf{Y}$ is entirely mediated through $\mathbf{X}$, or when their association with $\mathbf{Y}$ arises solely from common causes that are blocked by conditioning on $\mathbf{X}$. The modification to consider $\mathcal{G}_{\overline{\mathbf{Z}(\mathbf{X})}}$ rather than the original graph $\mathcal{G}$ accounts for the subtlety that we only consider intervening on components of $\mathbf{Z}$ that do not affect our conditioning set $\mathbf{X}$. This distinction is essential for properly capturing the notion of conditional causal irrelevance.
\end{proof}
\end{document}